# In-Plane Anisotropy of Electrical Transport in $Y_{0.85}Tb_{0.15}Ba_2Cu_3O_{7-x}$ Films

**Matvey Lyatti**[1,2,*], **Ines Kraiem**[1,2], **Torsten Röper**[1,2], **Irina Gundareva**[1,2], **Gregor Mussler**[1,2], **Abdur Rehman Jalil**[3], **Detlev Grützmacher**[1,2], **Thomas Schäpers**[1,2]

1. Peter Grünberg Institut (PGI-9), Forschungszentrum Jülich, 52425 Jülich, Germany
2. JARA-Fundamentals of Future Information Technology, Jülich-Aachen Research Alliance, Forschungszentrum Jülich and RWTH Aachen University, Germany
3. Peter Grünberg Institut (PGI-10), Forschungszentrum Jülich, 52425 Jülich, Germany
* Correspondence: m.lyatti@fz-juelich.de

**Abstract:** We fabricate high-quality c-axis oriented epitaxial $YBa_2Cu_3O_{7-x}$ films with 15% of yttrium atoms replaced by terbium (YTBCO) and study their electrical properties. The Tb substitution reduces the charge carrier density resulting in increased resistivity and decreased critical current density compared to pure $YBa_2Cu_3O_{7-x}$ films. The electrical properties of the YTBCO films show an in-plane anisotropy in both the superconducting and normal state that together with XRD data provide evidence for, at least, a partially twin-free film. Unexpectedly, the resistive transition of the bridges also demonstrates the in-plane anisotropy that can be explained within the framework of Tinkham's model of the resistive transition and the Berezinskii–Kosterlitz–Thouless (BKT) model depending on the sample parameters. Measurements of the differential resistance in the temperature range of the resistive transition confirm the occurrence of the BKT transition in the YTBCO bridges. We consider YTBCO films to be a promising platform for both the fabrication of devices with high kinetic inductance and the fundamental research on the BKT transition in the cuprate superconductors.

**Keywords:** high-temperature superconductivity; thin-film; in-plane anisotropy; superconducting transition.

## 1. Introduction

Superconducting films with a high kinetic inductance are considered a promising platform for many quantum applications including quantum computing and quantum communication. The kinetic inductance ($L_K$) arises from the kinetic energy stored in the motion of the charge carriers rather than the energy stored in the magnetic field. An impedance of nanoscale superconducting devices and Josephson junctions, where the kinetic inductance of the supercurrent dominates at operating frequencies, can achieve very high values. Nowadays, the high-kinetic inductance devices are based on disordered low-temperature superconductors such as NbN [1-3], NbTiN [4], NbSi [5], natural Josephson weak link arrays formed in the granular aluminum [6], artificial objects such as Josephson junction arrays [7,8], or semiconductor structures with induced superconductivity [9]. An alternative approach to the high-kinetic inductance materials for quantum applications could be based on high-temperature cuprate superconductors, where the fully gapped state is created by finite size effects, reduction of the doping level, or phase fluctuations [10-16]. Cuprate superconductors have low charge carrier density and, hence, high kinetic inductance per square

$$L_{k\square} = m_e/2de^2n_s = \hbar\rho_n/\Delta\pi d = \mu_0\lambda^2/d,$$

where $m_e$ is the free electron mass, $e$ is the electron charge, $n_s$ is the superfluid density, $\mu_0$ is the permeability of free space, $\hbar$ is the Plank constant, $\rho_n$ is the normal-state resistivity, $\lambda$ is the magnetic field penetration depth, and $d$ is the film thickness [17,18]. The in-plane magnetic penetration depth $\lambda$ of $YBa_2Cu_3O_{7-x}$ (YBCO) is comparable to the values reported for NbN films. Further increase in the kinetic inductance of YBCO films without creating oxygen vacancies is possible by the substitution of Y with another rare-earth metal [19,20]. Among different rare-earth metals, Tb is known for its ability to increase the YBCO in-plane resistivity by more than one order of magnitude without reduction of the critical temperature for a Tb content up to 50% [21-24]. Since the kinetic inductance is proportional to the resistivity, the very high resistivity might be a sign of a high kinetic inductance of the Tb-substituted YBCO films. However, the exact mechanism for the increase in the resistivity of YBCO with increasing terbium content is not clear. On the one hand, the resistivity increase was attributed to the Tb in the mixed-valence state that can reduce the hole doping [21]. On the other hand, the mixed-valence state of the Tb was not confirmed by the neutron-diffraction and x-ray data [25].

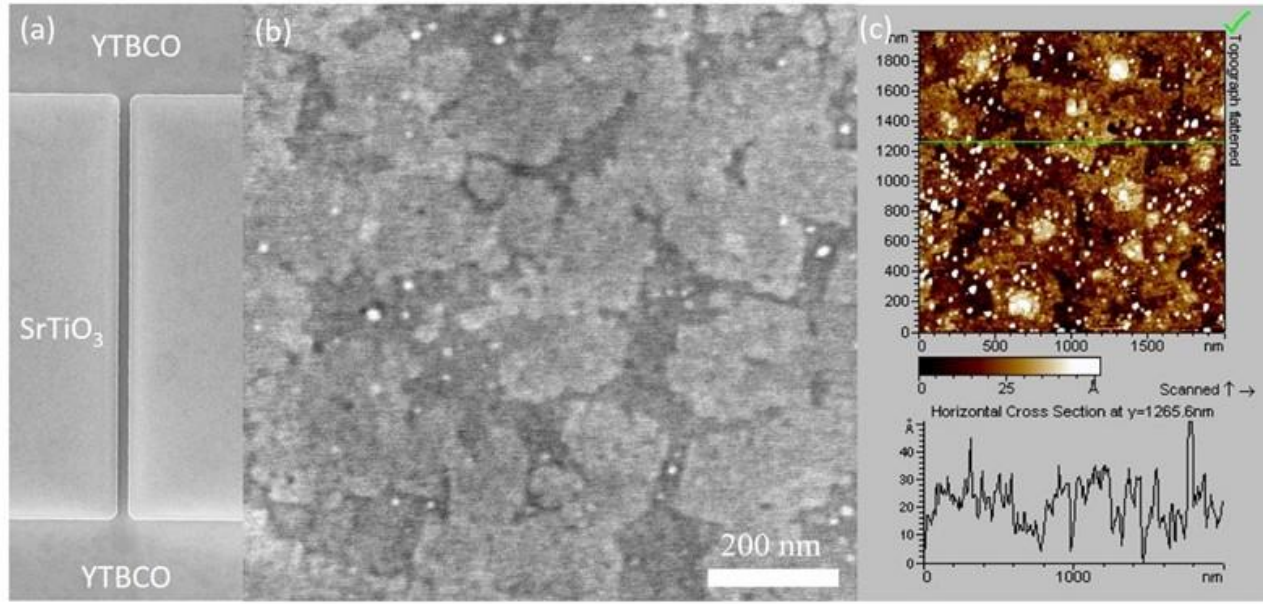

**Figure 1**. SEM micrograph of the 100-µm-long YTBCO bridge with the measured width of 3 µm (a). Images of the YTBCO film surface obtained with scanning electron (b) and atomic force microscopy (c).

We fabricated high-quality $Y_{0.85}Tb_{0.15}Ba_2Cu_3O_{7-x}$ (YTBCO) films and studied their electrical properties. We find that YTBCO films have in-plane anisotropy in both normal and superconducting states that together with XRD data provide evidence for the twin-free film. The in-plane anisotropy of the resistive transition can be explained within the framework of the Tinkham's model of the resistive transition [26] and the Berezinskii–Kosterlitz–Thouless (BKT) theory depending on the sample parameters [27]. Measuring the Hall effect in YTBCO as well as in pure YBCO films, we show that the increased resistivity of the YTBCO films is due to the decrease of the charge carrier density.

## 2. Experimental Details

We deposit the epitaxial YBCO films with 15 % of yttrium atoms replaced with terbium on (100) $SrTiO_3$ substrates by dc sputtering of a $Y_{0.85}Tb_{0.15}Ba_2Cu_3O_{7-x}$ target at a high oxygen pressure of $p(O_2) = 3.4$ mbar. Substrate edges are aligned along (010) and (001) planes. The temperature of the substrate heater is 935°C. The surface of the (100) $SrTiO_3$ substrates has a $TiO_2$ surface termination by etching in a buffered oxide etch. The film thickness $d$ is controlled by the sputtering time. The deposition rate of 1.38 nm/min is calibrated with X-ray reflectometry (XRR) and profilometer measurements of the film thickness. After the film deposition, the heater temperature is ramped down to 500°C with a rate of 30°C/min and then the deposited films are annealed during 15-30 min in pure oxygen at the pressure of 800 mbar. Following the annealing, the film is cooled down to room temperature at the rate of 30°C/min. The films thinner than 21 nm are additionally protected by a 7 nm thick nonsuperconducting amorphous YTBCO layer deposited *in situ* at 50-60°C heater temperature which is required to protect the films during structuring. The oxygen doping in films with the amorphous YTBCO layer on top may be lower than that in films without a protective layer due to plasma heating during the deposition of the amorphous layer. We did not perform additional oxygen annealing to avoid increasing the surface roughness of the amorphous layer. In addition to the YTBCO films, we deposit reference pure YBCO films with the same technique for the Hall effect measurements.

The YTBCO films are patterned into long bridges oriented along the [100] and [010] substrate crystallographic axes by a contact UV lithography and a chemical wet etching in a Br-ethanol solution. The bridges patterned in the PMMA resist along one crystallographic direction have a width $W$ of 3, 5, 10, and 20 µm and along another crystallographic direction a width of 5, 10, and 20 µm with a length $L$ to width ratio of $L/W = 20$. The YTBCO bridges are by 1-2 µm narrower than the corresponding PMMA resist patterns because of the undercut during the wet chemical etching. The actual bridge widths are determined by SEM measurements. The representative scanning electron microscopy (SEM) micrograph of the 3-µm-wide and 100-µm-long bridge is shown in Figure 1a.

Current-voltage (*IV*) characteristics of the current-biased bridges are measured at a zero magnetic field by a four-probe technique using self-made low-noise battery-driven electronics. The temperature dependence of the bridge resistance was obtained with a lock-in amplifier at a 10 kHz modulation frequency. The Hall measurements were performed at DC magnetic fields up to 1 T using current and magnetic field reversal.

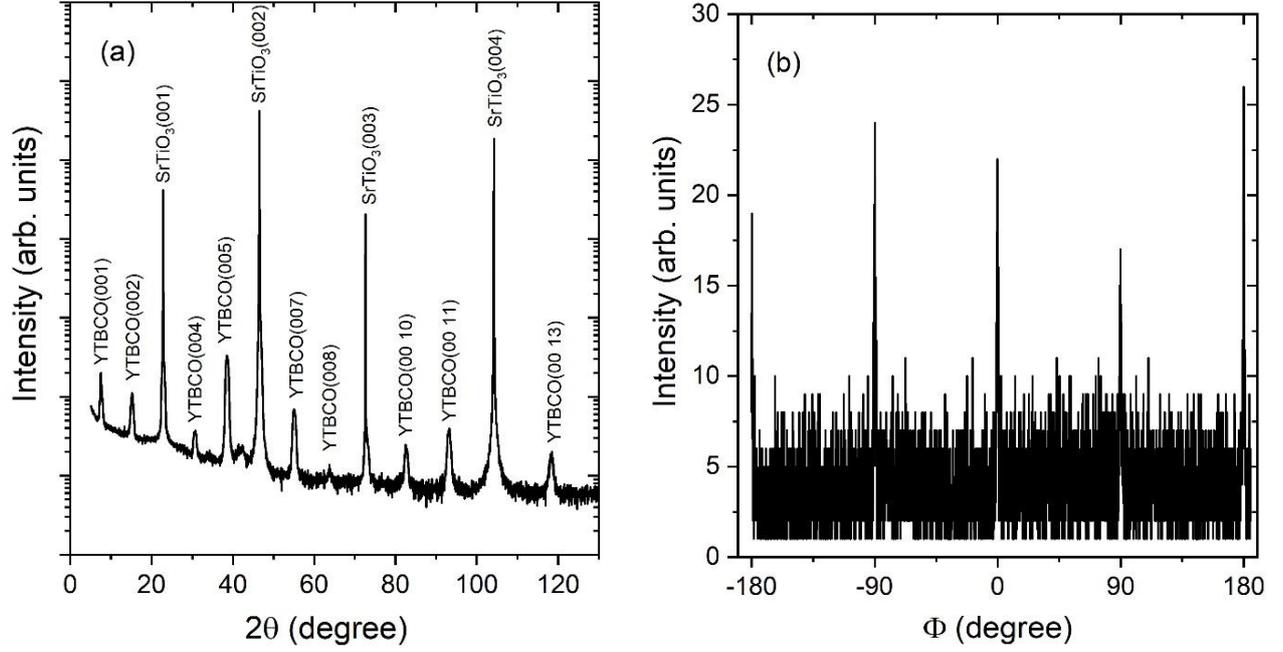

**Figure 2.** XRD 2Θ-Θ (a) and Φ (b) scans of a 35-nm-thick YTBCO film (sample N5).

## 3. Experimental Results and Discussion

### 3.1. Structural Analysis

We fabricate several YTBCO films with a thickness in the 11.8 – 123 nm range on the $SrTiO_3$ substrates. The representative images of the film surface obtained with SEM and an atomic force microscope (AFM) are shown in Figure 1b and Figure 1c. The films demonstrate layer-by-layer growth similar to those achieved for YBCO films in our previous work [28]. The contrast between the darker and lighter areas in the SEM image corresponds to the thickness difference of one unit cell (u.c.). The AFM micrograph shows that the surface of the YTBCO films has a root-mean-square roughness of 1.04 nm with round-shaped nano-precipitates with a diameter of 10-20 nm and a height of a few nm.

Since it has been shown that the formation of secondary phases is possible when the proportion of terbium exceeds 30% [22,29], we investigate the in-plane and out-of-plane crystallographic properties of our YTBCO films by x-ray diffraction (XRD) analysis with a high-resolution Bruker D8 Discover diffractometer. A 2Θ-Θ scan of the 35- nm-thick YTBCO film is shown in Figure 2a. Only (00l) YTBCO reflections are observed in the 2Θ-Θ scan, indicating single-crystalline growth with the *c*-axis in the growth direction. No secondary phase is seen. From the angular position of the (00l) peaks, we calculate the length of the YTBCO *c*-axis lattice parameter as 11.65 Å. The *a*- and *b*-axis lattice parameters are 3.86 Å and 3.87 Å – measured by means of reciprocal space maps (RSM) around the (1,0,8) and (0,1,8) reflections using a high-resolution Rigaku Smartlab diffractometer. The RSM cross sections presented in Figure 3 show two distinct peaks related to the *a*- and *b*-axis lattice parameters of YTBCO film. The small difference between the *a*- and *b*-axis lattice parameters resulting in overlapping RSM peaks does not allow us to draw a quantitative conclusion about the degree of the film twinning based only on the XRD data. We examined the films using polarized light optical microscopy and found no evidence of twins. We attribute the reduced difference between *a*- and *b*-axis lattice parameters compared to pure YBCO samples to the effect of Tb substitution [30]. The in-plane crystallographic structure is further analyzed by the Φ-scan. The Φ-scan around (104) and (014) reflections is plotted in Figure 2b. The peaks spaced with a periodicity of 90° for both reflections demonstrate good in-plane order of the film.

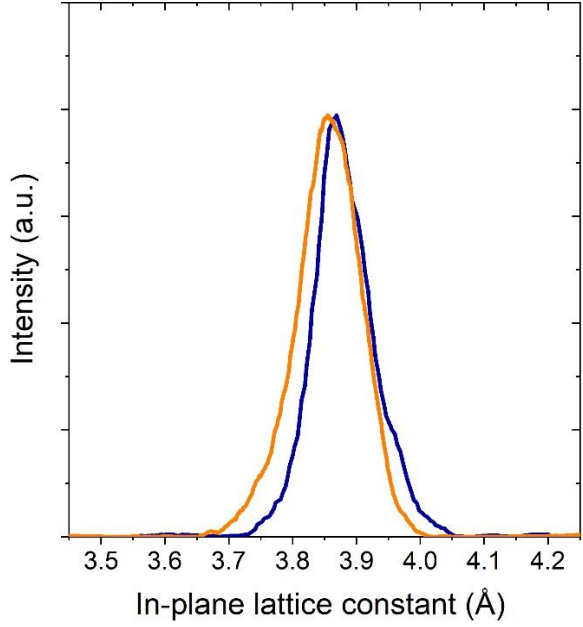 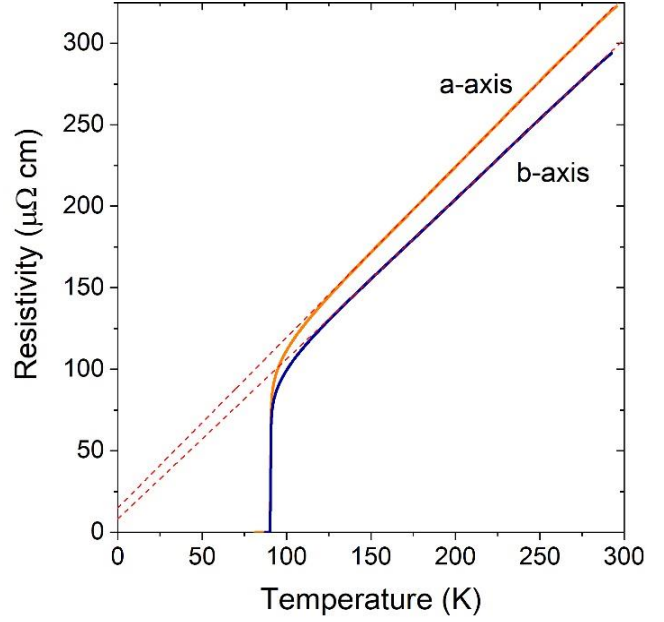

**Figure 3.** Cross sections of reciprocal space maps (RSM) around the (1,0,8) (orange curve) and (0,1,8) (blue curve) reflections.

**Figure 4.** Resistivity temperature dependence of the 8-μm-wide and 123-nm-thick bridges oriented in a and b crystallographic directions (sample N7). Dashed lines represent linear fits.

*3.2. Electrical Transport in normal state*

The normal-state resistance of the YTBCO bridges has a metallic temperature dependence with a sharp superconducting transition as observed for the pure YBCO films [28]. Taking into account the nonconducting layers at the top and bottom YTBCO film interfaces [28], we calculate the resistivity as

$$\rho_n = R[W \cdot (d-2\text{ u.c.})]/L,$$

where $R$ is the bridge resistance. The resistivity of the YTBCO bridges oriented along different substrate edges demonstrate clear in-plane anisotropy of the resistivity, as shown in Figure 4. Here we assume that the bridges with higher resistivity are aligned in the ***a***-axis direction and the bridges with lower resistivity in the ***b***-axis direction as it is observed for the pure YBCO [31]. The results of the resistivity calculations are presented in Table 1.

We find that the thicker films have a $\rho_n(300K)/\rho_n(100K)$ ratio close to 3 which is typical for the optimally doped YBCO films while this ratio for the thinner films with the protection layer is $\rho_n(300K)/\rho_n(100K) = 2.4\text{-}2.7$ which is a signature of underdoped films. The details of the $\rho_n(300K)/\rho_n(100K)$ ratio calculations are discussed in the Supplementary Materials. The lower oxygen concentration in the thinner films is likely due to the plasma heating during the protection layer deposition.

To calculate the in-plane resistivity anisotropy ratio, we use the resistivity values at $T = 100$ K. We find the in-plane anisotropy of the resistivity in the range of 1.05-1.13 for the optimally doped films and 1.08-1.45 for the underdoped films, as shown in Table 1. The measured in-plane anisotropy is smaller than that of the twin-free YBCO samples where it is 1.6-2.2 [30,34]. The reduced in-plane anisotropy may be related to the reduced difference between ***a***- and ***b***-axis lattice parameters due to the terbium substitution [35].

The resistivity of optimally doped YTBCO films is twice higher than the resistivity of the optimally-doped YBCO films deposited with the same technique [28]. To find out the reason for the increased resistivity of the YTBCO films, we measure the Hall coefficients and the Hall mobility for the YTBCO films at room temperature. Since the charge carrier concentration can depend on the deposition technique and oxygen annealing conditions, we also fabricated a pure YBCO reference film using the same sputtering and annealing parameters. The measurements were performed with eight-terminal Hall bars and Hall crosses oriented in the **a** and **b** crystallographic directions. The reference 30-nm-thick YBCO film shows the room-temperature resistivity of 208±3 μΩ·cm and the Hall coefficient of $(7.2\pm0.1)\cdot10^{-4}$ cm$^3$/C. The YBCO film doesn't demonstrate the in-plane anisotropy due to film twinning. The Hall coefficient of the YBCO films is close to the values reported for optimally doped YBCO single crystals and films [32,33]. In contrast to the twinned YBCO film, both the Hall coefficients and the resistivity of the YTBCO films have an in-plane anisotropy which is typical for twin-free YBCO samples [34]. Two representative YTBCO films with the average room-temperature resistivity of 307±13 μΩ·cm and 419±19 μΩ·cm show the average Hall coefficients of (9.45±0.05)·10⁻³ cm³/C and (1.31±0.06)·10⁻³ cm³/C, respectively. Therefore, we conclude that the Tb substitution reduces the charge carrier concentration in the YTBCO as it is expected in the case of the mixed-valence state of the Tb [21]. The smaller charge carrier concentration favors larger kinetic inductance which is the goal of this work.

**Table 1.** Parameters of the YTBCO samples including the film thickness $d$, the midpoint critical temperature $T_{c,mid}$, the superconducting transition width $\Delta T_c$, the critical current density $J_{ca}$ and $J_{cb}$, and the normal-state resistivity $\rho_{na}$ and $\rho_{nb}$ in $a$- and $b$-axis directions, respectively

| N | d [nm] | $T_{c,mid}$ [K] | $\Delta T_c$ [K] | $J_{ca}$(77.4K) $J_{cb}$(77.4K) [MA/cm²] | $J_{cb}/J_{ca}$ | $\rho_{na}$(100K) $\rho_{nb}$(100K) [μΩ·cm] | $\rho_{na}/\rho_{nb}$ | $\rho_n$(300K)/ $\rho_n$(100K) |
|---|---|---|---|---|---|---|---|---|
| 1 | 11.8 | 87.7 | 1.9 | 1.25±0.04 / 2.15±0.13 | 1.72 | 111.3±0.3 / 103±1.06 | 1.08 | 2.4±0.1 |
| 2 | 17 | 88.7 | 1.7 | 1.25±0.05 / 2.48±0.09 | 1.98 | 177±5 / 122±7 | 1.45 | 2.70±0.03 |
| 3 | 21 | 90.1 | 2 | 5.09±0.08 / 5.97±0.09 | 1.17 | 91.5±0.5 / 81±2.4 | 1.13 | 2.92±0.09 |
| 4 | 28 | 88.6 | 1.3 | 5.09±0.15 / 6.07±0.18 | 1.19 | 91.5±0.4 / 85±2.4 | 1.08 | 3.02±0.03 |
| 5 | 35 | 89.7 | 0.7 | 5.41±0.10 / 5.79±0.27 | 1.07 | 89.3±0.1 / 85.3±0.1 | 1.05 | 3.03±0.02 |
| 6 | 123 | 90.7 | 0.4 | 4.32±0.03 / 5.32±0.09 | 1.23 | 109±3.5 / 99±0.4 | 1.10 | 2.99±0.04 |

*3.3. Electrical Transport in superconducting state*

The transport properties of the YTBCO bridges in the superconducting state are studied by measuring *IV* curves of the bridges at $T$ = 77.4 K. The *IV* curves demonstrate a typical flux-flow behavior in the resistive state. We determine the critical current values $I_c$ with the 10 μV criterion and calculate the critical current density as $J_c = I_c/[W\cdot(d-3.\text{u.c.})]$ taking into account the nonsuperconducting layer at the YTBCO-substrate interface [28]. The calculated critical current densities show clear in-plane anisotropy with the $J_{cb}/J_{ca}$ ratio in the range of 1.07 – 1.23 for the optimally doped films and 1.72 – 1.98 for the underdoped films where $J_{ca}$ and $J_{cb}$ are critical current densities in the *a*- and *b*-axis crystallographic directions, respectively. The corresponding values are presented in Table 1. To confirm the reliability of the critical current density measurements, we fabricate an additional sample where seven bridges of the same width are oriented in the same direction. The 2% standard deviation of the critical current values for these bridges is well below the measured in-plane anisotropy of the critical current density. Therefore, we conclude that the difference between $J_c$ values in the *a*- and *b*-axis directions is due to the anisotropy of the film properties. The in-anisotropy of the critical current density of the YTBCO films is close to that reported for partially [35] and completely twin-free YBCO samples [36,37]. We believe that the in-plane anisotropy of both normal-state and superconducting properties of YTBCO films together with XRD and optical microscopy data provide evidence that our YTBCO films are at least partially untwinned.

Both the critical current density and the conductivity of the optimally-doped YTBCO films are approximately two times lower than those for the YBCO films fabricated with the same technique. Therefore, we expect that the kinetic inductance of the YBa₂Cu₃O₇₋ₓ films, where 15% of yttrium atoms are replaced with Tb, is twice as high as in pure YBCO films. A further increase of the kinetic inductance is possible at higher Tb concentrations.

*3.4. In-Plane Anisotropy of Superconducting Transition*

The average midpoint critical temperature $T_{c,mid}$ of the YTBCO films, corresponding to the middle of the transition, and the superconducting transition width $\Delta T_c$ are in the range of 87.7-90.1 K and 0.4-2 K, respectively, as shown in Table 1. These values are very close to those for pure YBCO films of the same thickness [28]. The superconducting transition width was determined according to a 90-10% resistivity drop criterion. The narrow superconducting transition, which is only 0.15 K broader than the best values reported for the YBCO single crystal samples, confirms the high quality of the YTBCO films [38].

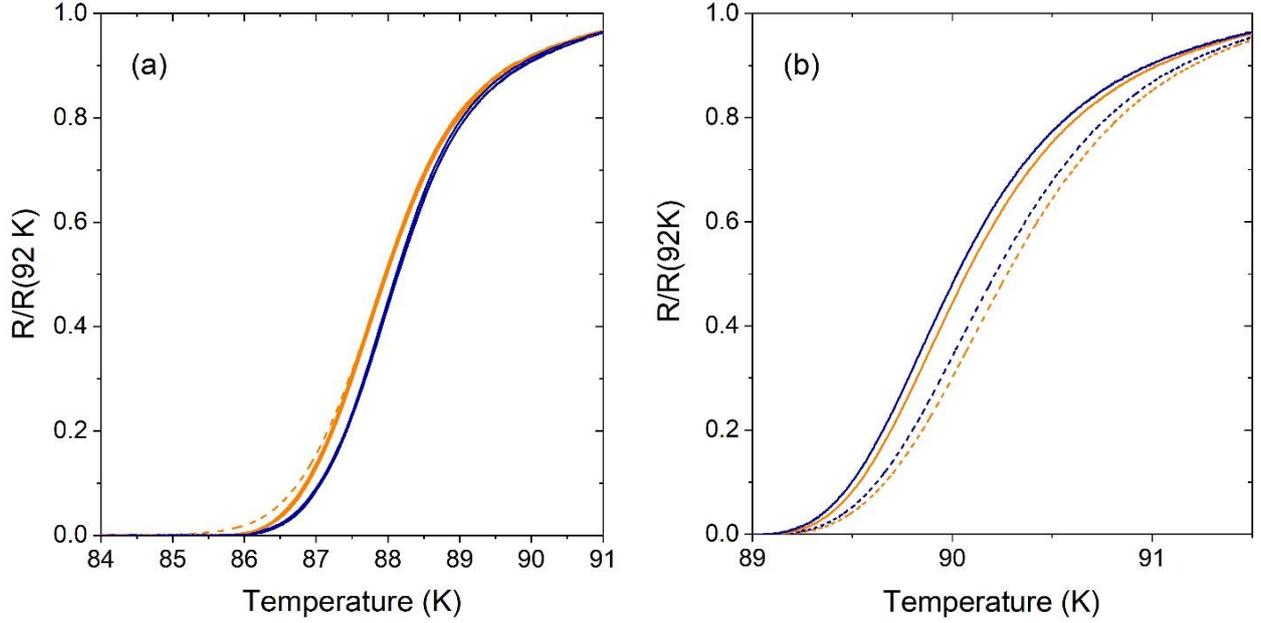

**Figure 5.** (a) The superconducting transition of seven 11.8-nm-thick YTBCO bridges (sample N1) with the width ranging from 2 to 19 μm oriented along the **a** (orange lines) and **b** (blue lines) crystallographic axes. The $R(T)$ curve of the narrowest bridge is shown by the dashed line. (b) The superconducting transition of four 21-nm-thick YTBCO bridges (sample N3) with the width of 19 μm (solid line) and 9 μm (dashed line). The bridges oriented along the **a** and **b** crystallographic axes are shown in orange and blue, respectively.

Measuring the $R(T)$ dependences, we noticed that the bridges of the same width patterned on the same substrate but oriented along different crystallographic axes have slightly different midpoint critical temperatures which might be a signature of the film inhomogeneity or related to the in-plane anisotropy. To find out the origin of the critical temperature scattering, we did a more thorough study of the superconducting transition ramping the temperature up and down in the 75 - 95 K temperature range. We gradually decrease the temperature ramp rate until the temperature hysteresis of the $R(T)$ dependence becomes less than the difference between the critical temperature of the bridges oriented along different crystallographic axes. Also, we reduce the ac bias current until it has no measurable influence on the critical temperature value. The final $R(T)$ dependence curve is calculated as an average of the $\varrho_n(T)$ curves measured during the ramping temperature up and down. Representative $R(T)$ dependences acquired for the seven 11.8-nm-thick bridges with widths ranging from 2 to 19 μm on the same substrate are presented in Figure 5a. The seven $R(T)$ dependencies are divided into two groups. Within each group, $R(T)$ dependencies match perfectly. The $R(T)$ dependencies with higher $T_c$, shown in blue, correspond to the bridges with higher critical current density, while the $R(T)$ dependencies with lower $T_c$, shown in orange, correspond to the bridges with lower critical current density. Since the bridges with higher critical current density have lower normal-state resistivity, they are oriented in the **b**-axis direction, while the bridges with lower critical current density are oriented in the **a**-axis direction. The higher resistance of the narrowest 2-μm-wide bridge (dashed orange line) at low temperatures may be due to the thermally activated phase slips or finite size effects [39]. Therefore, we conclude that the difference in $T_c$ values is because of the in-plane anisotropy rather than the film inhomogeneity. Remarkably, samples N3 and N4 show opposite behavior when the bridges with the higher critical current densities have a lower critical temperature, as shown in Figure 5b.

Since the anisotropy observed in the region of the phase fluctuations is caused by thermally generated vortices, we consider below two possible explanations of the superconducting transition anisotropy. The first one is the anisotropy of a vortex motion and the second one is an anisotropic Berezinskii–Kosterlitz–Thouless transition.

The description of the resistive transition in high-$T_c$ superconductors is quite complicated [40]. One of the simple explanations of the resistive transition in the high-$T_c$ superconductor is a thermally activated vortex motion. Within the framework of Tinkham's model of the resistive transition, a flux-flow resistance can be found as

$$R_{ff} = R_n [I_0(U_0/2k_B T)]^{-2},$$

where $R_n$ is normal state resistance, $U_0$ is the activation energy, $k_B$ is the Boltzmann constant, and $I_0$ is the modified Bessel function [26]. We find that our experimental $R(T)$ curves are in good agreement with the predictions of Tinkham's model as shown in Supplementary Materials. The activation energy anisotropy should result in the anisotropy of the flux-flow resistance. The higher critical temperature is expected for the bridges with the higher critical current density because of the activation energy $U_0 \sim J_c$ [26]. The resistive transition anisotropy of the thinnest samples N1 and N2 and the thickest samples N5 and N6 is in good agreement with Tinkham's model. In addition, the in-plane anisotropy of the vortex flow may also originate from the anisotropic distribution of the lattice defects such as grain boundaries.

However, samples N3 and N4 demonstrate a reversed sign of the resistive transition anisotropy when the bridges with higher critical current density have a lower critical temperature. This observation obviously cannot be explained by the in-plane anisotropy of the activation energy or the grain-boundary orientation because it does not depend on the film thickness. The reversed anisotropy of the resistive transition of the long anisotropic bridges may be due to the BKT transition.

The BKT transition is the transition between a low-temperature phase with logarithmically interacting vortices that form bound states at low temperatures and a high-temperature phase where the thermal fluctuation disrupts the bound states at the temperature above the BKT transition. Within the framework of the BKT model, all thermally generated vortices are paired below the BKT transition temperature $T_{BKT}$ resulting in a zero resistance. The observation of the BKT transition is expected in a 2D superconductor with a size $l$ larger than the effective magnetic penetration depth in a thin film

$$\Lambda = 2\lambda(T)^2/d,$$

where $\lambda$ is the magnetic penetration depth [41]. If the size of the sample is significantly smaller than $\Lambda$, vortex-antivortex pairs are not formed [42]. At the intermediate size of the sample, the vortex-antivortex pairs and free vortices may coexist below $T_{BKT}$ [43], and we assume that the concentration of unpaired vortices and thus the resistance increase at $T < T_{BKT}$ with the decrease of the $l/\Lambda$ ratio.

In the case of long bridges, the BKT transition precondition is fulfilled in the longitudinal direction $L > 2\lambda(T)^2/d$ but may be broken in the transversal direction depending on the bridge width. Then, the resistive transition of the long bridges with the intermediate width is controlled by the $W/\Lambda_t$ ratio where $\Lambda_t$ is the transversal effective magnetic penetration depth. Bridges of the same width and thickness but with different magnetic penetration depths in the transverse direction have different resistances at the same temperature within the range of the resistive transition.

Our anisotropic YTBCO bridges oriented in the *a*- and *b*-axis direction have critical current densities $J_{ca} < J_{cb}$ and $\rho_{ca} > \rho_{cb}$. Based on that we can expect the magnetic penetration depths $\lambda_a > \lambda_b$, and, hence, the $W/\Lambda_t$ ratio in the transverse direction $W/\Lambda_b > W/\Lambda_a$, respectively, for the same bridge width W. Then in the case of the BKT transition, the YTBCO bridges with the lower critical current density have to have a higher critical temperature compared to those with higher critical current density because they have the larger $W/\Lambda_t$ ratio in the transverse direction resulting in the reversed resistive transition anisotropy. Therefore, an observation of the reversed in-plane anisotropy of the resistive transition in the anisotropic YTBCO films may be a new signature of the BKT transition in the cuprate superconductors.

The observation of the BKT transition in the cuprate superconductors is very challenging. While the broadening of the superconducting transition and the decrease in the zero-resistance temperature with the decrease of the film thickness, and the kinetic inductance jump predicted by the BKT model have been consistently observed [43-47], evidence of the jump in the exponent $\alpha$ is inconsistent [48-52]. Inconsistency of the experimental data led to the discussion on the existence of the BKT transition in the superconducting films. It was assumed that the vortex pinning [53] or finite-size effects can mimic the BTK behavior [54,55] or destroy it [42]. It was argued that the attempts to observe BKT transition could be unsuccessful due to inhomogeneity, improper sample size, or current noise [56]. Our YTBCO films have a smaller difference between *a* and *b*-axis parameters compared to pure YBCO films, which should lead to a smaller number of defects induced by the tetragonal–orthorhombic phase transition. Therefore, the reversed resistive transition anisotropy may indeed be due to the BKT transition.

To confirm the existence of the BKT transition in the studied YTBCO bridges, we examine a hallmark of the BKT transition which is the change of the exponent $\alpha$ in the current-voltage characteristic $V \sim I^\alpha$ from one to three at the temperature above and below the BKT transition, respectively [27]. The nonlinear behavior of the superconducting film at $T < T_{BKT}$ is due to the current-induced unbinding of vortex-antivortex pairs. We measure the differential resistance $R_d = dV/dI$ of the bridges with an intermediate film thickness, where the occurrence of the BKT transition is suspected. Figure 6a shows a representative series of $R_d(I)$-$R_d(0)$ dependences of the 19-μm-wide and 30-nm-thick bridge in the 88 - 90 K temperature range. The $R_d(I)$ dependences are measured within the ± 9 μA current range to avoid the influence of the bias current on the resistive transition. The differential resistance demonstrates the current-independent behavior at temperatures above 88.215 K and parabolic behavior at temperatures below 88.915 K which is a clear sign of the current-induced unbinding of vortex-antivortex pairs below the BKT transition temperature. The zoomed $R_d(I)$ curves at the temperatures above and below the BKT temperature and the results of experimental curves fitting by the $R_d$ = const and parabolic dependencies are shown in the inserts in Figure 6b by black and red lines, respectively. The transition between current independent and parabolic $R_d(I)$ dependence is very sharp with a width below 20 mK. The temperature dependence of the $α$-1 value and the resistive transition are shown in Figure 6b. Orange points indicate the temperatures where $R_d(I)$ curves are measured. The vortex-antivortex pairing occurs at the temperature above the zero-resistance temperature that rules out the non-linearity caused by the vortex dynamics at currents above the critical current value. The BKT transition in the studied YTBCO bridges is not complete. Free-vortices and vortex-antivortex pairs coexist below the BKT temperature which might be due to finite-size effects. In the case of the coexistence of the vortex-antivortex pair and free vortices, only the $α$-1 exponent in the $dV/dI \sim I^{α-1}$ dependence shows the jump from 0 to 2 below the BKT temperature, as observed for our YTBCO bridges, while the change of the $α$ exponent in the $V \sim I^α$ dependence is continuous.

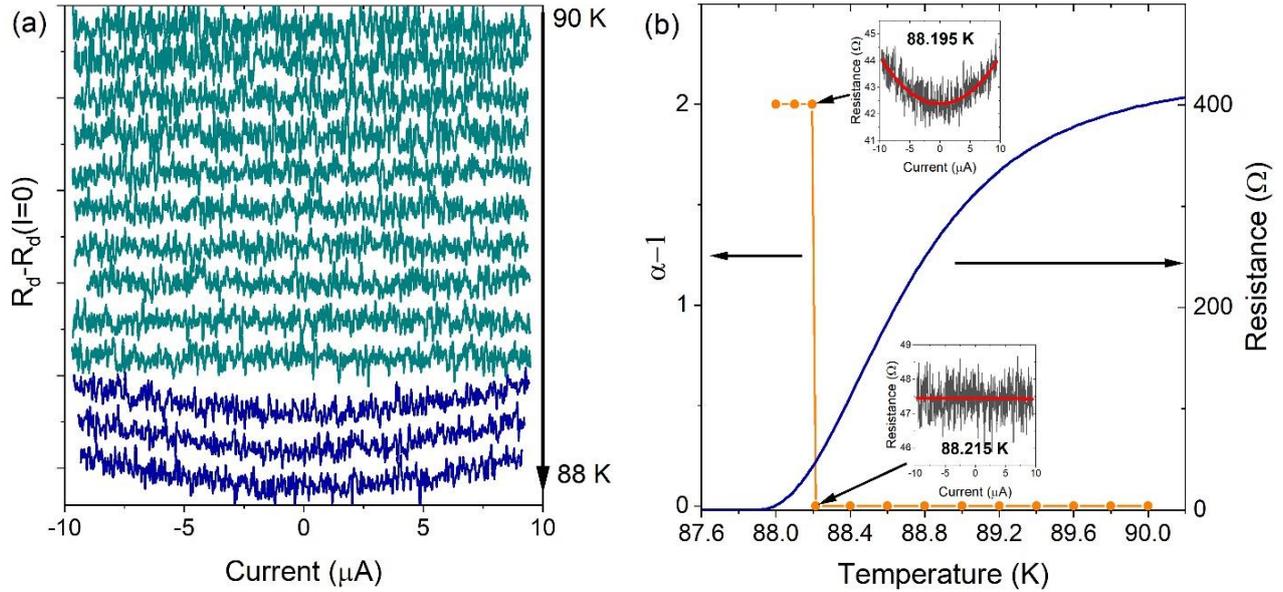

**Figure 6.** (a) The $R_d(I)$-$R_d(I=0)$ dependences of 19-μm-wide and 30-nm-thick YTBCO bridge in the 88 – 90 K temperature range. Curves are shifted along the ordinate axis for the convenience of their representation. Curves above and below the BKT transition are shown in dark cyan and dark blue, respectively. (b) The temperature dependences of the 19-μm-wide and 30-nm-thick YTBCO bridge resistance (dark blue line) and the $α$-1 exponent value (orange line). Inserts show $R_d(I)$ dependences at $T$ = 88.195 and 88.215 K.

## 4. Conclusions

In conclusion, we fabricate high-quality YTBCO films with a very low surface roughness that exhibit the in-plane anisotropy in both superconducting and normal states. We find that the increased resistivity and reduced critical current density are due to the reduced charge carrier density in YTBCO films compared to YBCO films which is beneficial for high-kinetic inductance devices. We observe the sharp switching of the $R_d(I)$ dependence of the YTBCO bridges in the resistive transition region from current independent to the parabolic behavior which we consider as a clear sign of the BKT transition. The reversed in-plane anisotropy of the resistive transition observed in the long YTBCO bridges can be explained within the framework of the BKT transition. We consider the YTBCO films to be promising for the fabrication of devices with high kinetic inductance and the investigation of the BKT transition in cuprate superconductors.

# SUPPLEMENTARY MATERIALS for

# In-Plane Anisotropy of Electrical Transport in $Y_{0.85}Tb_{0.15}Ba_2Cu_3O_{7-x}$ Films

Matvey Lyatti[1,2*], Ines Kraiem[1,2], Torsten Röper[1,2], Irina Gundareva[1,2], Gregor Mussler[1,2], Detlev Grützmacher[1,2], Thomas Schäpers[1,2]

1 Peter Grünberg Institut (PGI-9), Forschungszentrum Jülich, 52425 Jülich, Germany

2 JARA-Fundamentals of Future Information Technology, Jülich-Aachen Research Alliance, Forschungszentrum Jülich and RWTH Aachen University, Germany

* Correspondence: m.lyatti@fz-juelich.de

**Temperature dependence of the YTBCO film resistance**

The $\rho_n(T)$ of the thick films follows linear temperature dependence (Fig. S1a) while the $\rho_n(T)$ of the thinnest films has a significant deviation from the linear dependence as shown in Fig. S1b. This deviation from the linear dependence becomes more pronounced when the $Y_{0.85}Tb_{0.15}Ba_2Cu_3O_{7-x}$ (YTBCO) film thickness is reduced to a few unit cells. We attribute the deviation of the $\rho_n(T)$ dependence from the linear dependence observed for the thinnest films to the contribution of the few unit cell thick non-superconducting layer at the film-substrate interface that has a semiconducting type of conductivity. The semiconducting layer contribution distorts the ratio between the resistance values at 100 and 300 K $R(300K)/R(100K)$ which is frequently used to characterize the quality and the doping level of the superconducting cuprate films. To obtain the contribution of the superconducting layer with the metallic type of the conductivity, we model the resistance of the film at the temperature above the critical temperature by a parallel connection of the resistors with the metallic ($A + BT$) and semiconducting $C \cdot \exp(E_a/2k_BT)$ types of the conductivity, where $A$, $B$, $C$, and $E_a$ are constants, $T$ is the temperature, and $k_B$ is the Boltzmann constant. This simple model fits experimental curves well in the 120-300 K temperature range. At temperatures below 120 K, the experimental curve lies below the fitting curve because of the Aslamazov-Larkin fluctuation conductivity that is not included in the model. The $R(T)$ dependence with and without semiconducting layer contribution and the semiconducting layer contribution are shown in Fig. S1b by black, red, and blue lines respectively. To calculate the $R(300K)/R(100K)$, we use the $R(T)$ dependences with subtracted semiconducting layer contribution. For the same reasons, we determine the in-plane anisotropy of the film resistivity at $T$ = 100 K where the contribution of the semiconducting layer is negligible.

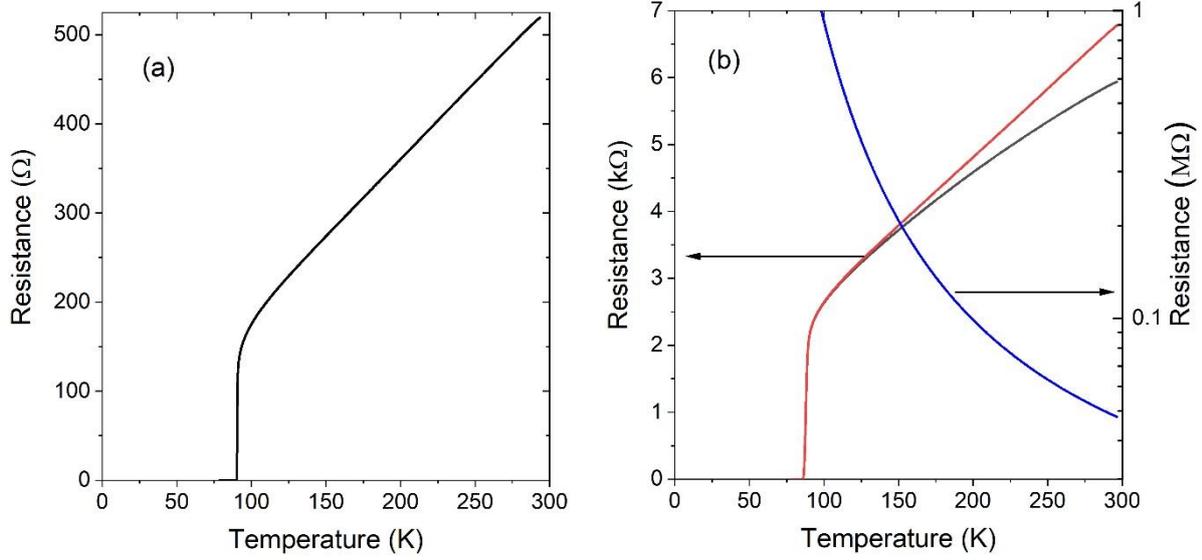

FIG. S1. Resistance temperature dependence of the 126-nm-thick (a) and 11.8-nm-thick (b) YTBCO bridges. The measured *R(T)* dependence, the *R(T)* dependence with the subtracted contribution of the semiconducting layer, and the semiconducting layer contribution are shown by black, red, and blue lines, respectively.

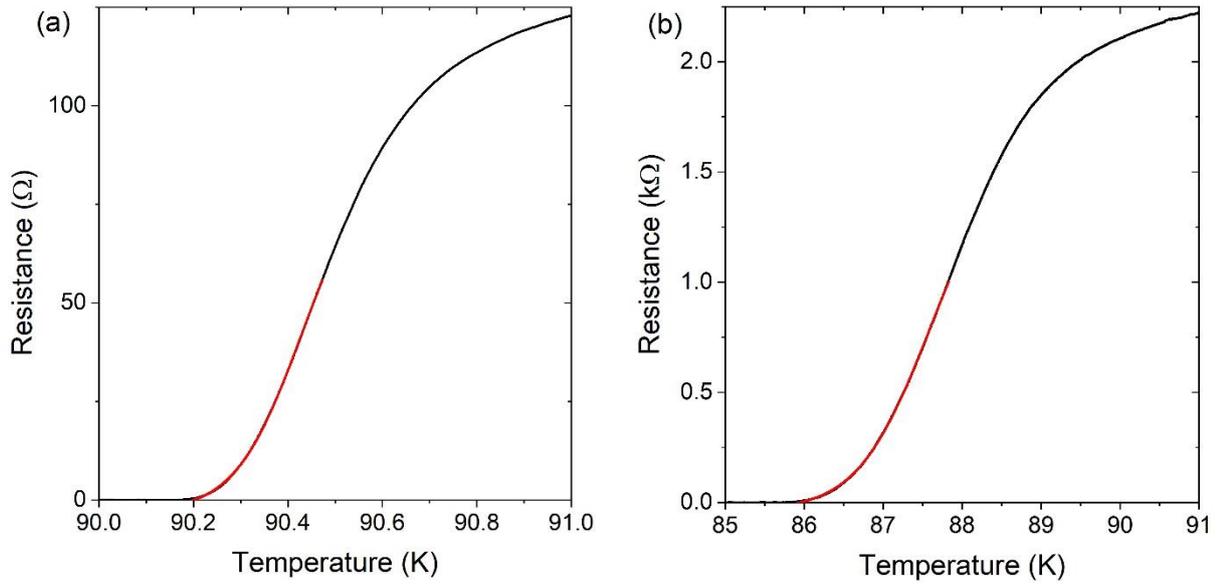

FIG. S2. Fitting of the resistance temperature dependence of the 19-μm-wide YTBCO bridges with the thickness of (a) 123 nm and (b) 11.8 nm (black lines) by the Tinkham's model (red lines).